\documentclass[journal=jacsat,manuscript=article]{achemso}
\SectionNumbersOn
\usepackage[version=3]{mhchem} 
\usepackage{hyperref}
\hypersetup{colorlinks=true,citecolor=blue,linkcolor=red,urlcolor=blue}
\usepackage{placeins}
\usepackage{color}
\usepackage{amsmath,amssymb}

\definecolor{green}{rgb}{0,0.5,0}

\author{Melisa M. Gianetti}
\affiliation{Dipartimento di Fisica, Università degli Studi di Milano, Via Celoria 16, Milano 20133, Italy}
\altaffiliation{Current affiliation: Institutt for maskinteknikk og produksjon, NTNU, Richard Birkelands vei 2B, 7034 Trondheim, Norway}
\email{melisamariel@gmail.com}

\author{Roberto Guerra}
\affiliation{Center for Complexity and Biosystems, Department of Physics, University of Milan, via Celoria 16, Milano 20133, Italy}

\author{Andrea Vanossi}
\affiliation{CNR-IOM,
Consiglio Nazionale delle Ricerche - Istituto Officina dei Materiali and International School for Advanced Studies (SISSA), Via Bonomea 265, 34136 Trieste, Italy}

\author{Michael Urbakh}
\affiliation{Department of Physical Chemistry, School of Chemistry, The Raymond and Beverly Sackler Faculty of Exact Sciences and The Sackler Center for Computational Molecular and Materials Science, Tel Aviv University, Tel Aviv 6997801, Israel}

\author{Nicola Manini}
\affiliation{Dipartimento di Fisica, Università degli Studi di Milano, Via Celoria 16, Milano 20133, Italy}

\title{Electric-field frictional effects in confined zwitterionic molecules}


\begin{document}

\begin{tocentry} 

\includegraphics[width=0.6\textwidth]{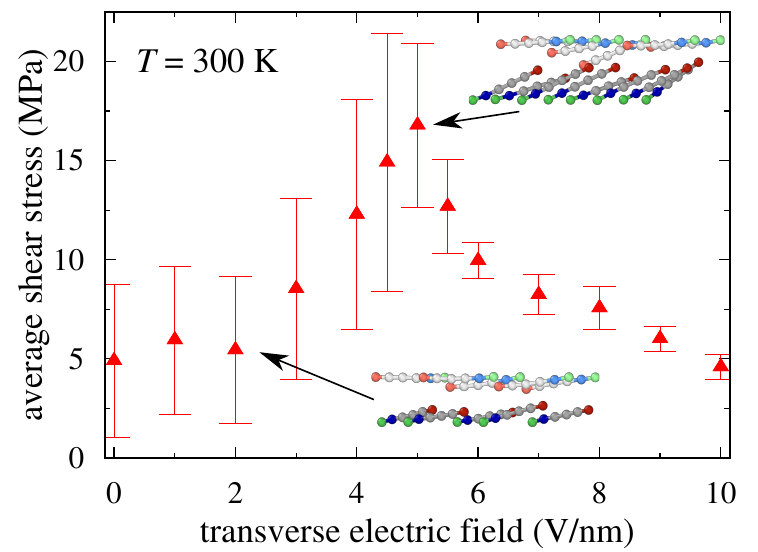}


\end{tocentry}

\begin{abstract} 
{
We theoretically explore the effect of a transverse electric field on the frictional response of a bi-layer of packed zwitterionic molecules. 
The dipole-moment reorientation promoted by the electric field can lead to either stick-slip or smooth sliding dynamics, 
with average shear stress values varying over a wide range.
 A structure-property relation is revealed by investigating the array of molecules and their mutual orientation and interlocking.
  Moreover, the thermal friction enhancement previously observed in these molecules 
  is shown to be suppressed 
  by the electric field, recovering the expected thermolubricity at large-enough fields. 
  The same holds for other basic tribological quantities, such as the external load, which can influence friction in opposite ways depending on the strength
  of the applied electric field. 
  Our findings open a route for the reversible control of friction forces via electric polarization of the sliding surface.
}
\end{abstract}

\section{Introduction}

Given the substantial interest for applications, the possibility of controlling friction and mechanical response without direct intervention on the often inaccessible contacting surfaces by means of external fields  
has been investigated extensively in recent years \cite{Krim19,Spikes2020,Bresme2022,Song2022}.
In particular, the tuning role of applied electric fields in confined geometries
has been widely investigated in triboelectrochemistry experiments \cite{Labuda2011,Kailer2011,sheng2012electrorheological,Hausen2013,Iqbal2015,Vanossi18,pashazanusi2018,Ma2021,Li2022}.
In most of these experiments ions move in a liquid solution 
driven by the
applied field, covering and effectively modifying the sliding surfaces, thus inducing changes in friction 
\cite{Fedorov2008,Li2014ionic,Yang2014,Capozza15a,Fajardo15b,Capozza15b,Dold2015,Strelcov2015,Pivnic2018}.
An alternative approach is based on the ability of the electric field to reorient macromolecules, thus changing their conformation in aqueous solution 
\cite{Lahann2003,Drummond2012,Zeng2017} 
and dry environments \cite{karuppiah2009,deWijn2014,deWijn16}, with potentially dramatic effects on friction. 
Within this alternative scheme, here we consider neutral chain molecules tethered to 
flat 
substrate surfaces. 
These chains represent 
zwitterionic molecules, which have found applications for colloid stabilization, regulation in wetting and adhesion, the creation of protective coatings and many others
\cite{Lahann2003,myshkin2009adhesion,ma2019brushing,qu2022}. 
Zwitterionic molecules can also be reoriented by using light \cite{dubner2014light} and in some particular cases the 
orientation of these large molecules or some parts of them 
can be determined using X-ray spectroscopy \cite{hahner1996self}. 
The characterization of the structure of polymer layers deposited on surfaces is always crucial for tribology \cite{gellman2022}. 
The 
surface force apparatus (SFA) 
provides a highly sensitive way to measure the frictional shear stress between atomically flat surfaces with molecules deposited on them, while applying a controlled normal load \cite{Drummond2000,Drummond01,Drummond03,chen2009,raviv2003}.
While such experimental setup does measure the system rheological and dissipative response in terms of crucial, yet averaged, physical quantities \cite{Carpick97}, a viable exploitation of the electrotunable approach requires
casting light on the elemental mechanisms and molecular rearrangements occurring at the sheared interface.
Molecular dynamics (MD) simulations, 
as a sort of controlled computational ``experiment'', have proved to be extremely useful in investigating the atomistic details of frictional processes at sliding interfaces \cite{Szlufarska08,VanossiRMP13,Manini15,Manini16,Monti2020,frerot2023}.
These strategies make it possible to avoid interpretative pitfalls arising from indirect or \textit{ex situ} characterization of contact surfaces.

This work explores
this idea, considering a model where the
electric-field-controlled
geometric rearrangement of 
zwitterionic molecular segments leads to 
modifications of the sliding interface properties, 
associated with distinct frictional regimes.

\section{\label{sec:methods}Model and Simulations}

\begin{figure}[ht]
\centering
\includegraphics[width=\columnwidth]{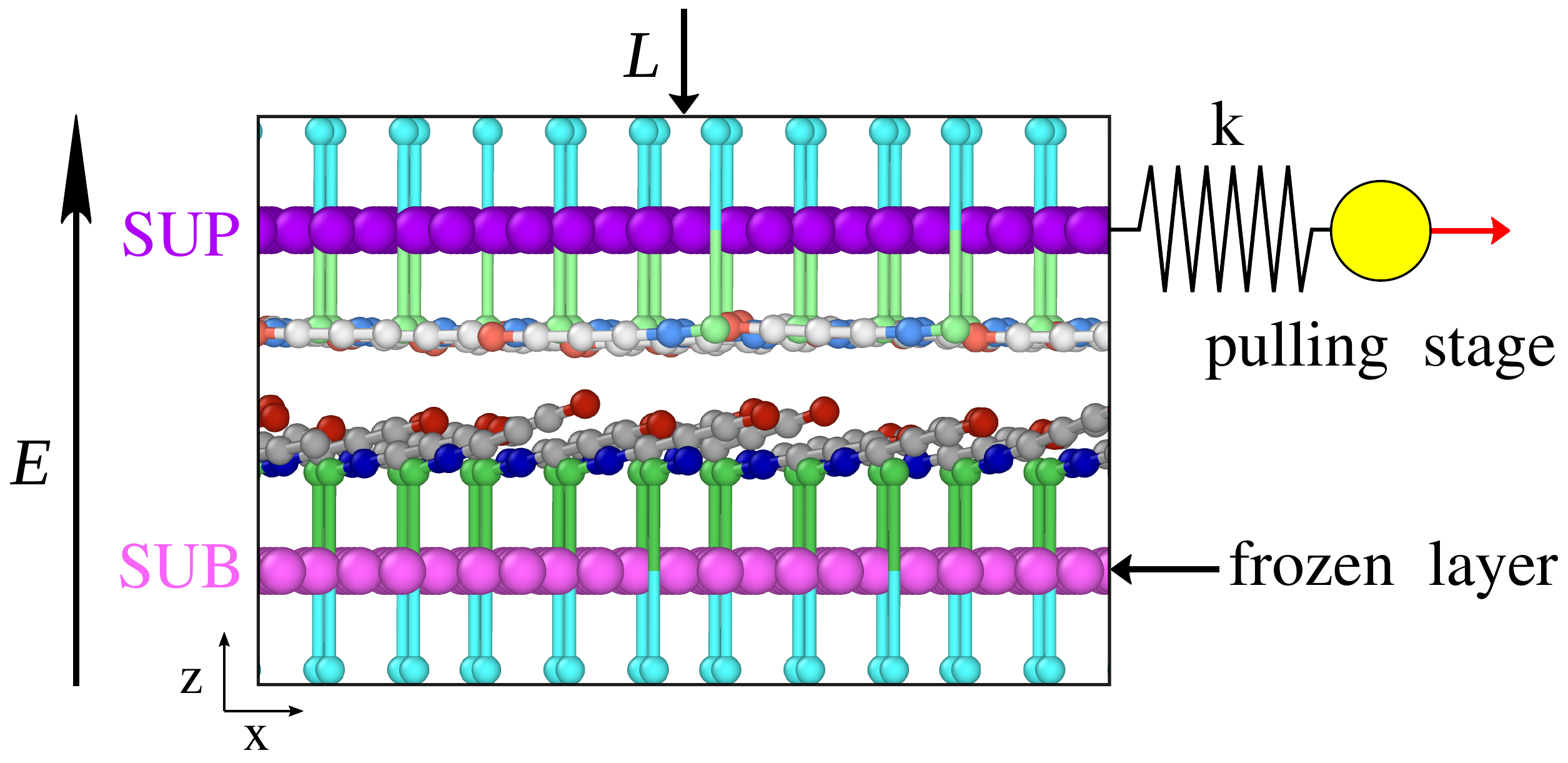}
\caption{
  Side view of a snapshot of the contacting zwitterionic molecules 
  at $T = 0~\text{K}$, with an applied electric field $E = 5~\text{V}\cdot\text{nm}^{-1}$ and load $L = 10~\text{MPa}$.
  Red, gray, blue, green, cyan spheres represent cations, neutral residues, anions and the two uncharged particles standing for a glycerol group and the hydrophobic chain forming the inner part of the vesicle, respectively. 
  For clarity, SUP chains are colored lighter than SUB chains. 
  Magenta and purple spheres represent the static SUB and the sliding SUP rigid layers, respectively.
  The yellow sphere represents the pulling stage advancing at constant speed and driving the SUP layer through a spring.
}
\label{fig:system}
\end{figure}

We extend a recently-developed model inspired by 
SFA
experiments with confined self-assembled vesicles of zwitterionic molec\-ules \cite{Klein2016,Klein2019Langmuir,lin2020cells}.
The model was introduced and described in detail in Ref.~\citenum{gianetti2022}. 
Briefly, the organic polymers composing the vesicles walls are represented as chains of point-like particles.
Each of these particles is a coarse-grained representation of 
the dipalmitoylphosphatidylcholine 
molecule, as detailed in Figure~1 of Ref.~\citenum{gianetti2022}, 
and as investigated in SFA experiments very recently 
with similar molecules.\cite{Zhang2023} 
As the protagonists of interfacial friction are the zwitterionic head groups, the molecular model describes them in greater detail.
Each chain of 7 point particles consists of: one cation followed by three uncharged residues, by an anion, and finally by two uncharged particles representing a glycerol group and a long alkyl tail (see Figure~\ref{fig:system}).
In real-life self-assembled vesicles, these hydrophobic tails segregate away from the aqueous solution, provide a directed support to the zwitterionic segments, and transmit the external load and shear forces produced by the SFA probe.
In the model, the structural hydrophobic parts of the vesicles 
are represented by parallel rigid layers, that can slide relative to each other, in a supercell of area $A \simeq 120~\text{nm}^2$. 
We name the 
sliding layer SUP and the static one SUB, see Figure~\ref{fig:system}.
The two end uncharged residues of each molecule are `planted' in these rigid layers, which keep them in place, while allowing them some elastic freedom.
To account for screening by the surrounding water molecules, the charges on the ionic residues are reduced to $\pm 25\%$ of an elementary charge.

Compared to Ref.~\citenum{gianetti2022},
the novelty of the present work is the introduction of an electric field directed along the $\hat z$ axis, i.e.\ perpendicular to the sliding, which affects the orientation of the zwitteronic molecules and their response to shear. 
In SFA experiments, such an electric field arises when the top surface is positively charged and the bottom one is negatively charged.
Because the total molecular charge vanishes, so does the net electric force acting on each molecule.
On the other hand, the zwitterionic section of the molecule carries an electric dipole given by the product of the residue charge, $0.25$ elementary charges, times their separation, which equals $0.64~\text{nm}$ at equilibrium.
Thus this dipole moment is approximately $d=2.6\times 10^{-29}~\text{C}\cdot\text{m} = 7.7~\text{Debye}$.
The electric field generates a torque acting on each molecular dipole.
This torque tends to deflect the zwitterionic chain orientation away from the equilibrium angle $\theta=111^\circ$ \cite{Bockmann2008kinetics,gianetti2022}
implemented in the adopted molecular model, toward the vertical direction $\hat z$.
The torque is maximum when the zwitterionic chain lies flat in the $xy$ plane ($\theta=90^\circ$), and vanishes when the chain stands upright in the $\hat z$ direction ($\theta=180^\circ$), which represents the stable equilibrium point for the electric torque produced by the upward electric field.
For the direction of the electric field considered here only the zwitterionic chains planted in the SUB layer can reach this ``standup'' orientation in the field, because for the molecules planted in the SUP layer any upward bending is hindered by the rigid SUP layer itself (see Figure~\ref{fig:system}).
As a result, an increasing strength of the electric field favors the SUB-related zwitterionic chains tendency to stand up, while it tends to push the upper layer of zwitterionic chains flatter and flatter in the $xy$ plane.
The electric-energy gain of a dipole rotating from the horizontal to the standup configuration in a $E = 1~\text{V}\cdot\text{nm}^{-1}$ field amounts to $E\cdot d=0.16~\text{eV}$.
For comparison, the angular-spring energy cost to rotate from the rest position at $111^\circ$ to vertical ($180^\circ$) amounts to
$1.45~\text{eV}$.

We adopt LAMMPS \cite{lammps,thompson2022lammps} 
as the simulation platform and we control the simulation temperature by means of a Langevin thermostat with a damping rate $\gamma=1~\text{ps}^{-1}$ applied to all particles forming the molecules \cite{Allen91}.
This thermostat is set to act only along the transverse coordinates
$y$ and $z$, in order to prevent any spurious thermostat-originated frictional damping along the most relevant sliding direction $
x$ \cite{Robbins01,rottler2003growth}.

We start off with a fresh initial configuration consisting in a periodic, but arbitrary chain arrays, as described in Ref.~\citenum{gianetti2022}.
To prepare a thermodynamically meaningful initial point, we set $E = 5~\text{V}\cdot\text{nm}^{-1}$ 
and a constant sliding velocity of $5~\text{m}\cdot\text{s}^{-1}$.
Under these conditions, we anneal the system by ramping $T$ up from 0 to 500~K in 1~ns, we keep it at 500~K for 1~ns, we then ramp the temperature down from 500 to 0~K, in 1~ns, and finally continue sliding at 0~K for 1~ns. 
The final configuration of this simulation is the starting point for multiple simulations with different values of $E$. 
Before each run, we ramp the electric field steadily up or down to the target field value $E$ in $100$~ps. 

Starting from the end of the $5~\text{V}\cdot\text{nm}^{-1}$ simulation at $T=0$, we ramp up the temperature to $300~\text{K}$ in 2~ns and then maintain it constant for another 0.2~ns to produce the initial configuration for all the $T = 300$\,K simulations. Different $E$ values are obtained with the same ramps as for the $T = 0$ case.
In a typical production run, we
pull 
the SUP layer through a spring with $k = 1~\text{eV}\cdot\text{nm}^{-2} \simeq  0.16~\text{N}\cdot\text{m}^{-1}$, and let the stage advance by 100~nm, usually at $v_\text{stage} = 5~\text{m}\cdot \text{s}^{-1}$, thus running for 20~ns.
Batches of simulations with varied loads are performed in sequences first increasing and then decreasing the load in small steps. 
The instantaneous friction force (and therefore the shear stress) is provided by the spring elongation.
Average quantities are evaluated over the final $80$~nm of 
steady-state regime, i.e.\ omitting the initial transient of $20$~nm.

\section{Results and discussion}

\subsection{Effects of the transverse electric field}
\begin{figure*}[ht]
\centering
\includegraphics[width=0.9\textwidth]{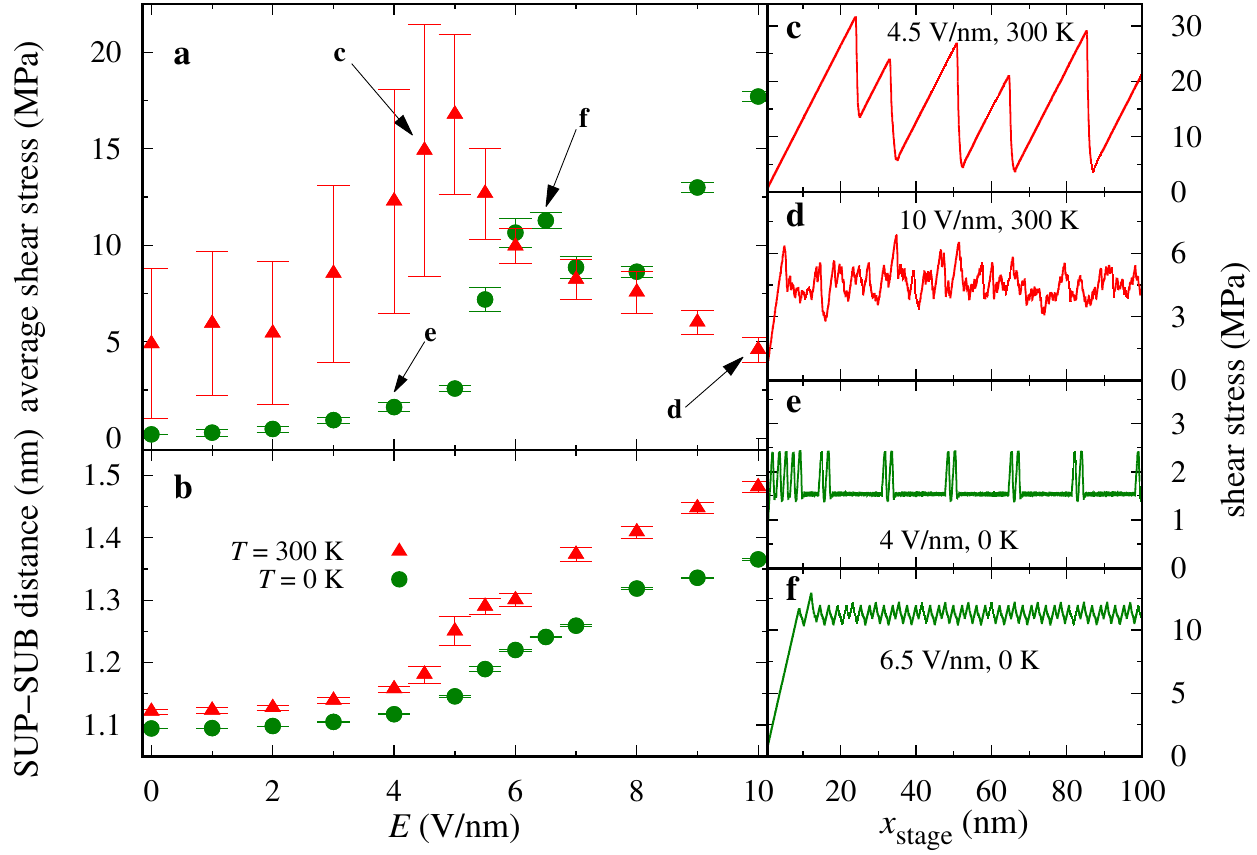}
\caption{
  Variation of (a) the average frictional shear stress and (b) the average distance between the rigid layers as a function of the electric field $E$, for $v_\text{stage} = 5~\text{m}\cdot \text{s}^{-1}$, applied load $L = 10~\text{MPa}$, and at temperatures $T = 0~\text{K}$ (green circles) and $T = 300~\text{K}$ (red triangles).
  Averaging omits the initial transient consisting in the first 20~nm of sliding in each simulation.
  Vertical 
  bars measure the root mean square fluctuations: wide bars are typically indicative of stick-slip dynamics.
  (c-f) Examples of friction traces (instantaneous shear stress as a function of $x_\text{stage}=v_\text{stage}\, t$), corresponding to the $E$ values pointed at in panel (a).
}
\label{fig:main}
\end{figure*}

\begin{figure*}[ht]
\centering
\includegraphics[width=\textwidth]{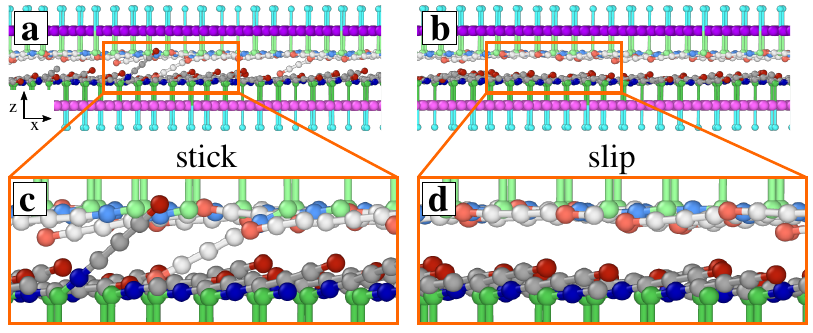}
\caption{\label{snaps300_low}
 5~nm $y$-thick slices of snapshots of a simulation carried out at $T = 300~\text{K}$, with $E = 2~\text{V}\cdot\text{nm}^{-1}$ and $v_\text{stage}= 5~\text{m}\cdot\text{s}^{-1}$. 
  Side (a,b) and zoomed (c,d) views for (a,c) stick and (b,d) slip configurations. 
  For clarity, SUP chains are colored lighter than SUB chains. 
}
\end{figure*}

Figure~\ref{fig:main}a reports the average frictional shear stress as a function of the electric field.
The vertical bars reflect the root mean squared fluctuations: wide bars are indicative of stick-slip dynamics, as, e.g., in the friction trace of Figure~\ref{fig:main}c, while smoother sliding generates narrower bars, as in the friction trace of Figure~\ref{fig:main}f.
The $T = 300~\text{K}$ simulation data are consistent with rather large ratio $S/L$ of shear stress to load in the 0.5--1.8 range, depending on the electric field.
The average friction is enhanced by the electric field up to $E\simeq 5~\text{V}\cdot\text{nm}^{-1}$ showing stick-slip dynamics for all electric-field values, as in the example of Figure~\ref{fig:main}c.
The 
upward pointing
electric field
flattens
the SUP chains
against the supporting layer: 
intra-layer steric hindrance and the Coulombic interactions 
within the same layer
generate a
substantially flat well-ordered configuration.
In contrast, the zwitterionic heads of the SUB chains tend to align in the electric field's direction, 
pointing in
a more and more vertical configuration as the electric field increases. 
The lifted heads acquire an increasing configuration freedom, which allows the cations to interact more effectively with the SUP chains: more and more of their head cations reach energetically convenient ``hollow sites'' 
formed by 
adjacent anions 
in the SUP layer 
(see Figure~\ref{snaps300_low}c). 
In the slip events this interlocking is lost (see Figure~\ref{snaps300_low}b,d and ESI Movie1~\dag). 
This increased interlocking tends to favor stick-slip dynamics, as evident from the data of Figure~\ref{fig:main}. 
%

\begin{figure*}[ht]
\centering
\includegraphics[width=\textwidth]{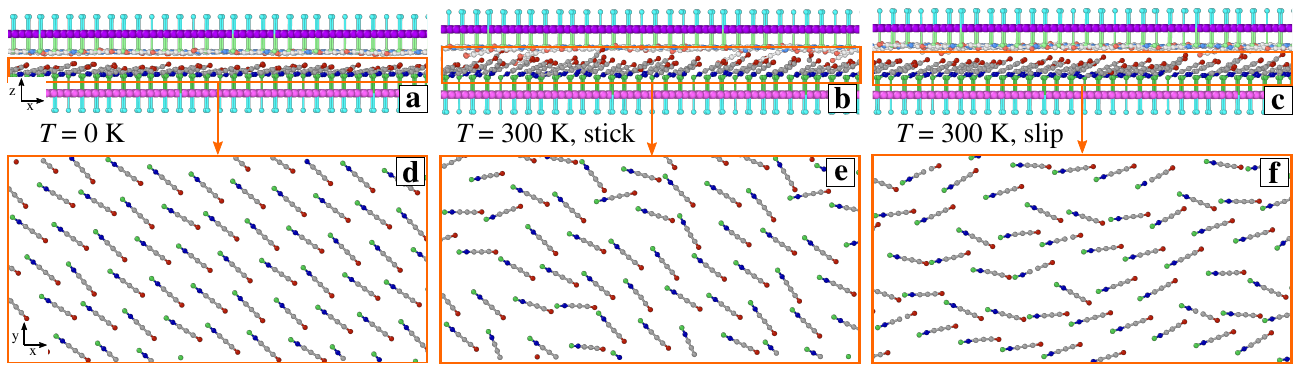}
\caption{
Side (a-c) and top (d-f) views of 4 nm $y$-thick slices of $E = 5~\text{V}\cdot\text{nm}^{-1}$ simulation snapshots at (a,d) $T = 0$ and (b,c,e,f) $T = 300~\text{K}$. 
Top views show only SUB chains inside the orange rectangles in (a-c). 
For clarity, SUP chains are colored lighter than SUB chains. 
}
\label{snapsEF_5}
\end{figure*}

For even larger fields $E \geq 5~\text{V}\cdot\text{nm}^{-1}$
the SUB layer rids itself of its steric hindrance and, as a result, the chains in this layer acquire a substantially disordered and fluid-like arrangement
(see Electronic Supplementary Information (ESI) Figure~S1b and Movie3~\dag).
The SUB chains liftup is also illustrated in the snapshots of Figures~\ref{fig:system} and \ref{snapsEF_5}, in ESI Movie2~\dag, and more quantitatively by the thickening of the molecular layer reported in
Figure~\ref{fig:main}b.
The SUP layer becomes extremely flat, while the excessive freedom makes the
SUB chains highly susceptible to thermal fluctuations, which tend to weaken the transient interlocking bonds, 
leading to the reduction of friction.
The ensuing
erratic
dynamics appears
somewhere in between stick-slip and smooth sliding, see Figure~\ref{fig:main}d. 
Note that partial alignment of the SUB chains in Figure~\ref{snapsEF_5}f is the result of the ``combing'' produced by the rightward advancing SUP layer.

Given the nontrivial temperature effects that were observed in this model in the absence of the transverse electric field \cite{gianetti2022}, it is useful to compare the presented results with those of a similar investigation
carried out at $T = 0~\text{K}$ (green data in Figure~\ref{fig:main}). 
Up to $E \simeq 4~\text{V}\cdot\text{nm}^{-1}$, the system exhibits 
essentially smooth-sliding dynamics with weak stick points (see e.g.\ Figure~\ref{fig:main}e) resulting in fairly low overall friction.
We observe a ratio $S/L =0.16$ at $E = 4~\text{V}\cdot\text{nm}^{-1}$, and $S/L<0.1$ for $E \leq 3~\text{V}\cdot\text{nm}^{-1}$.
Like at $T=300~\text{K}$, as $E$ is increased the layer of the SUP chains flattens out,
while the SUB chains tend to stand up and approach the vertical direction of the electric field.
However, for a moderate field $E<5~\text{V}\cdot\text{nm}^{-1}$ without thermal fluctuations the SUB layer remains mostly ordered, with very few SUB cations attain the energetically favorable hollow configuration. 
The result is a quite weak overall corrugation which is reflected in the low friction. 

\begin{figure*}[ht]
\centering
\includegraphics[width=\textwidth]{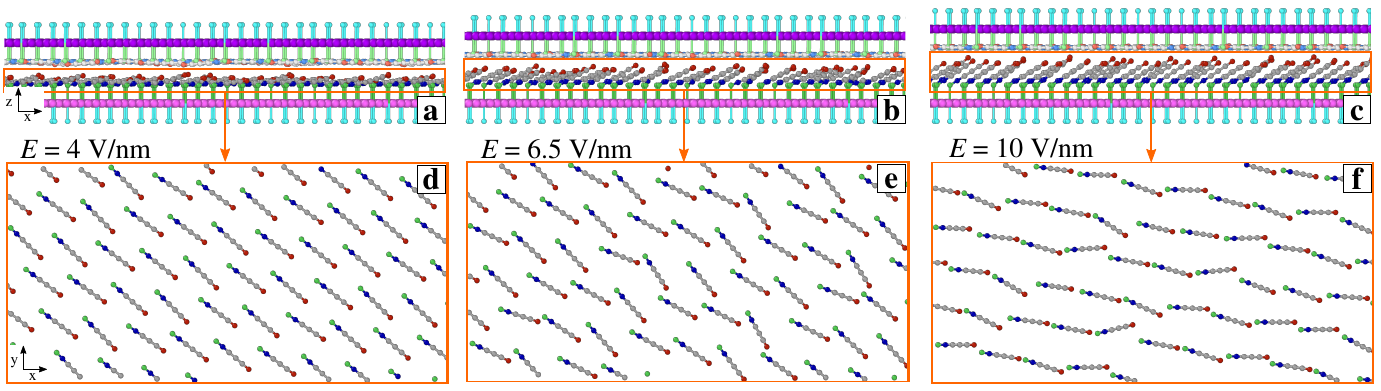}
\caption{
Side (a-c) and top (d-f) views of 4 nm $y$-thick slices of $T = 0~\text{K}$ simulation snapshots for different values of electric field: 
(a,d), (b,e), (c,f) correspond to 
$E = 4$, 6.5, and $10~\text{V}\cdot\text{nm}^{-1}$ 
respectively. 
Top views show only SUB chains inside the orange rectangles in (a-c). 
For clarity, SUP chains are colored lighter than SUB chains. 
}
\label{snaps0}
\end{figure*}

For $E \geq 5~\text{V}\cdot\text{nm}^{-1}$, 
friction increases rapidly because the SUB chains 
further stand up (see Figure~\ref{snaps0}) forming a looser and thicker layer, and acquiring such a substantial lateral freedom, that
a significant fraction of SUB cations succeeds in reaching the energetically favorable interaction sites surrounded by the anions of the flat SUP layer (like in Figure~\ref{snaps300_low}a).
The resulting extra SUB-SUP interaction enhances the effective corrugation, and therefore also friction at high field, see Figure~\ref{fig:main}a.

\subsection{Effects of sliding velocity}

\begin{figure*}[ht]
\centering
\includegraphics[width=0.9\textwidth]{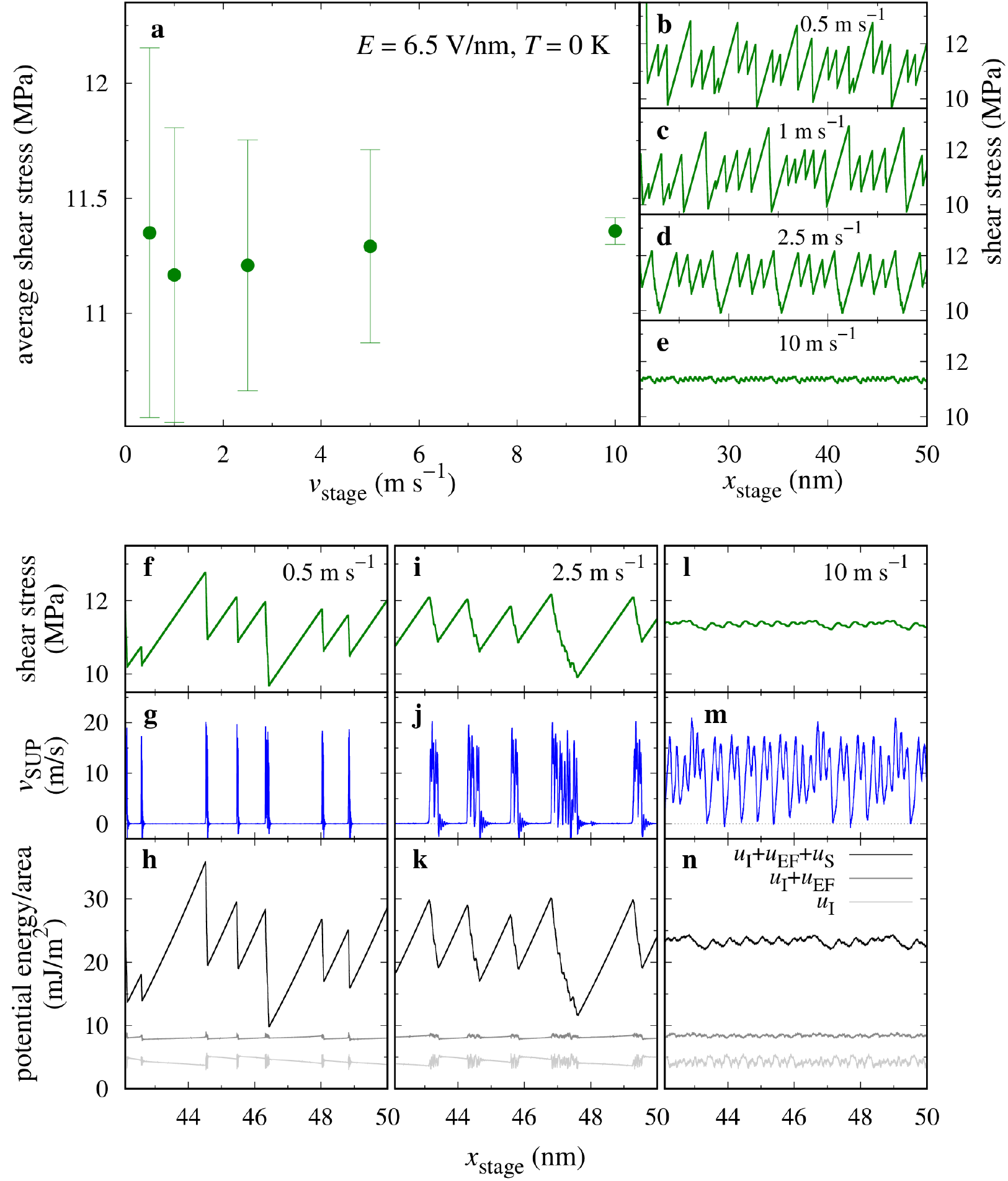}
\caption{
  (a) Average shear stress as a function of $v_\text{stage}$ for $E = 6.5~\text{V}\cdot\text{nm}^{-1}$, $T = 0~\text{K}$ and $L = 10~\text{MPa}$. 
  (b-e) Shear traces for a few of the velocities reported in panel (a). 
  (f,i,l) Shear stress,
  (g,j,m) instantaneous velocity of the SUP layer and 
  (h,k,n) per unit area potential energy contributions as a function of the stage displacement $x_\text{stage}$, for three of the explored velocities. 
  $u_\text{I} = U_\text{I}/A$ = internal potential energy;
  $u_\text{EF} = U_\text{EF}/A$ = potential energy of the charges in the applied electric field; 
  $u_\text{S} = U_\text{S}/A$ = potential energy of the pulling spring.
}
\label{fig:vX_0_6.5}
\end{figure*}

As the sliding speed is reduced, tribological systems often tend to
transition 
from smooth sliding to stick-slip dynamics.
The current model is expected to follow this tendency.
To verify this hypothesis, and to gain insight in the model behavior at lower speed, i.e.\ closer to an experimentally relevant regime, we tested the model 
behavior as a function of the velocity of the pulling stage 
for specific values of $T$ and $E$.
For $E = 6.5~\text{V}\cdot\text{nm}^{-1}$ at $T = 0~\text{K}$ the system
exhibits
stick-slip dynamics at most of the simulated velocities, see Figure~\ref{fig:vX_0_6.5}a-e.
This results in an 
essentially 
velocity-independent time-averaged friction, Figure~\ref{fig:vX_0_6.5}a.

The friction peaks acquire more and more the asymmetric shape, typical of stick-slip dynamics, as the driving speed is reduced.
The reason is evident by comparing the spikes in the velocity of the SUP center of mass as a function of the pulling stage displacement, reported in Figure~\ref{fig:vX_0_6.5}g,j,m. 
The slip jump, of the order of 1~nm, tends to increase marginally with the driving speed \cite{Labuda2011}. 
Similar features are observed for $E = 4~\text{V}\cdot\text{nm}^{-1}$ (see ESI Figure~S2~\dag).

To better characterize the origin of these distinct frictional regimes, 
we analyze the energetics of the system, considering the three contributions to the total potential energy: the internal potential energy ($U_\text{I}$) including the bonding, nonbonding, and Coulomb interactions within and among the molecules, the energy due to the interaction of the molecular charges with the applied electric field ($U_\text{EF}$) and that associated to the elongation of the pulling spring ($U_\text{S}$), see Figure~\ref{fig:vX_0_6.5}h,k,n, where combinations of these energies are reported per unit area, and shifted by irrelevant arbitrary constants.

Remarkably, the internal energy $U_\text{I}$ tends to decrease during the stick phase, showing that the molecular layers actually relax while the spring exerts a stronger and stronger pulling force.
Inclusion of the electrical energy causes the potential energy to slowly increase 
during stick, reflecting the increasing tilt of the SUB chains.
However, this progressive increase and the corresponding drop at slip are quite small, less than $0.5~\text{mJ}\cdot\text{m}^{-2}$, compared to the changes in the total potential energy, which includes the spring contribution too.
The smallness of these energy steps indicates an extremely small deformation and forward displacement of the SUP layer during the stick phase, with the slip event occurring as a sudden collective collapse of the
provisionally-formed 
interlayer bonds.
The total potential energy released in its downward jumps, of the order $1~\text{mJ}\cdot\text{m}^{-2}$, divided by the typical slip jump $\simeq 1~\text{nm}$ generates a typical stress in the $1~\text{MPa}$ region, quite consistent with the observed shear-stress peaks at the end of the stick intervals.

We verified (see ESI Figure~S3~\dag) that essentially the same observations regarding the effect of varying the driving velocity apply at room temperature, for a field in the same range,  $E=5~\text{V}\cdot\text{nm}^{-1}$.
As expected, the dynamics does not change for slower sliding velocity and the highest simulated velocity is not sufficiently large
to change the dynamics to smooth sliding.

Finally, for very large field $E=10~\text{V}\cdot\text{nm}^{-1}$, 
the dynamics is neither a clear stick-slip nor smooth a sliding. 
A speed reduction (as far as accessible in simulations) does not seem to lead
to a clearer
picture, see ESI Figure~S4~\dag for details.

%
%

\subsection{Effects of the applied load}

The applied load $L$ was 
kept at
$10~\text{MPa}$ in all previous simulations. 
To explore the effect of varying load, in simulations we raise and then decrease it in small steps.
As the friction traces show no significant memory effect, we report the resulting data as an average over both simulations executed under the same load.
We evaluate friction for a low ($1~\text{V}\cdot\text{nm}^{-1}$), intermediate ($5~\text{V}\cdot\text{nm}^{-1}$), and high ($10~\text{V}\cdot\text{nm}^{-1}$) electric field.
\begin{figure*}[ht]
\centering
\includegraphics[width=0.9\textwidth]{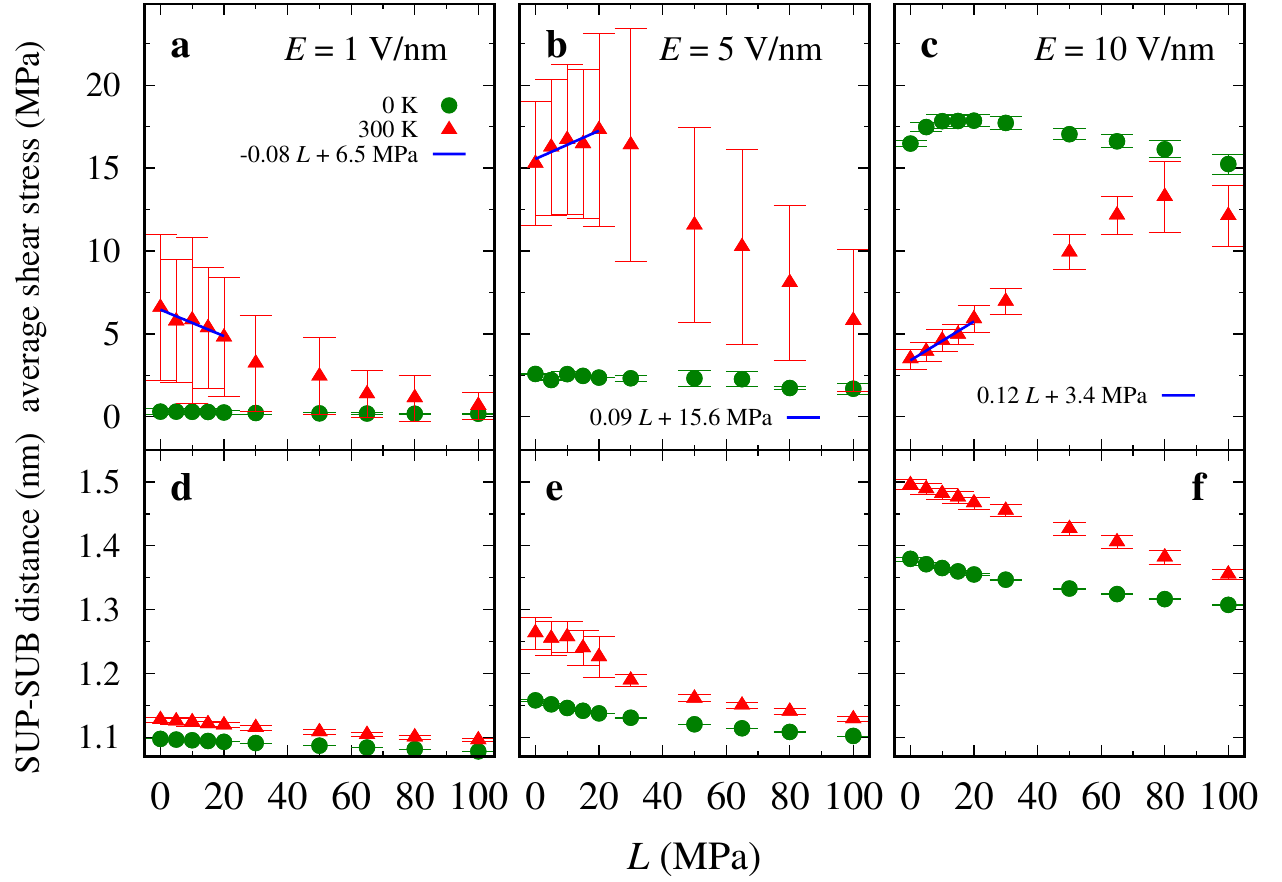}
\caption{\label{fig:loads}
(a-c) Average shear stress and 
(d-e) SUP-SUB distance as a function of the applied load $L$.
Blue lines are linear fits of the small-load (up to 20~MPa) $T=300~\text{K}$ data.
The resulting differential friction coefficients are:
$\mu = -0.08$ for $E =  1~\text{V}\cdot\text{nm}^{-1}$;
$\mu = 0.09$ for $E =  5~\text{V}\cdot\text{nm}^{-1}$;
$\mu = 0.12$ for $E = 10~\text{V}\cdot\text{nm}^{-1}$.
}
\end{figure*}

As shown in Figure~\ref{fig:loads}, 
the outcome of these simulations indicates that the load dependence of friction changes qualitatively, depending on the electric field.
In the absence of electric field, Ref.~\citenum{gianetti2022} reported an essentially load-independent frictional shear stress.
Here, for relatively weak field, we observe that friction decreases with load (differential friction coefficient $\mu \equiv dS/dL \simeq -0.08$).
For intermediate field, at room temperature the model exhibits a nonmonotonic dependence of friction upon loading, with friction initially increasing ($\mu \simeq 0.09$), and then rapidly decreasing at higher load.
At the highest tested field, friction increases faster ($\mu \simeq 0.12$) and the friction-growing range extends to even larger loads, although eventually friction reaches a maximum around $L\simeq 80~\text{MPa}$, and then it starts to decrease.
This
nontrivial
behavior emerges despite the expected monotonic compression of the sheared layer, shown in Figure~\ref{fig:loads}d,e,f. 

The mechanism for the friction change with load is related to the changes in the number of interpenetrating chains of opposite layers.
We quantify the degree of interlocking by evaluating the ``hooking fraction'' $h$, 
defined as 
the fractional number of chains whose cation crosses the average level of cations of the opposite layer \cite{gianetti2022}.

For low field, the increasing load promotes
steric
interactions 
among chains in the same layer,
resulting in a flatter configuration with 
suppressed
hooking (see Figure~S5a~\dag) and lower friction. 
Intermediate field pushes the SUB chains to a more vertical position.
Increasing load promotes interlocking, and thus increased friction, up to 20~MPa.
When the SUB chains bend substantially under the effect of higher loads, they acquire flatter layer configurations, and further raising $L$ suppresses interlocking (see Figure~S5~\dag), and therefore friction,
as observed in Figure~\ref{fig:loads}b. 
For large field the interlocking is negligible (less than 1\%) for all since 
the SUP chains adopt a substantially disordered configuration, as previously noted. 
Increasing load brings opposite chains closer, rapidly promoting stronger Coulombic and steric interactions, so that friction increases. 

For $T = 0~\text{K}$ load has generally moderate
effect on friction, with an overall tendency to suppress friction. 
The reason is the SUB layer being pressed down with decreased angular fluctuations of the individual chains.

\FloatBarrier

\section{Conclusions}

In mainstream triboelectrochemical approaches, the electric field introduces a bias to the diffusive ionic motion and
even moderate electric fields can generate a significant electric-energy difference for the ions at different locations, easily exceeding the thermal energy $k_\text{B} T$, thus resulting in significant ionic displacements, and potentially important surface alterations with effects on friction which can range from modest to dramatic.
The model investigated here involves no free ions: all frictional changes here are associated to the reorientation of the flexible dipolar section of zwitterionic molecules pinned to a surface. 
As a result, the displacement of charged residues is limited by the molecular size, and accordingly, for a given field strength, the reorientation energetics is comparably more modest than can be achieved by mobile ions.
Moreover, this electric-coupling energy has to compete not only with thermal disordering effects of the order of $k_\text{B} T$, but also with the elastic molecular deformation energy.
As a consequence, quite strong fields are generally required for sizable effects.
In the present model we consider fields up to $10~\text{V}\cdot\text{nm}^{-1}$, which determine substantial structural alterations and remarkable effects on friction.
Such field values can however be quite challenging to achieve in the lab.
With easily accessible fields much smaller than $1~\text{V}\cdot\text{nm}^{-1}$, we observe negligible structural and tribologic effects.
The dramatic tribologic effects observed very recently in phospholipid assemblies \cite{Jin2023Tunning} are likely associated to field-induced dramatic layer reconfigurations, such as the electroporation mechanism identified in that work and in Ref.~\citenum{Jin2023Electric-field}, which are clearly outside the globally rigid-layer model studied here.

The main outcome of this investigation is the non-monotonic variation of friction as a function of the electric field at room temperature. 
The electric field promotes an asymmetric deformation of the SUP and SUB zwitterionic layers. 
With increasing field the upper-layer chains tend to arrange in a compact ordered and flat configuration while the lower-layer chains tend to stand up and approach the vertical direction of the electric field.
Thermal fluctuations promote interlayer chain interlocking and the increased freedom to move of the standing-up SUB chains  allows a substantial fraction of them to reach binding sites in the SUP layer. 
This is reflected by an increase in friction for $E \leq 5.5~\text{V}\cdot\text{nm}^{-1}$. 
Higher fields lead to an extra-flat, compressed, and rigid SUP monolayer
that suppresses interlocking and thus friction. 

By testing the dependence on the sliding velocity for specific values of $T$ and $E$ we find that 
for moderate fields stick-slip dynamics is observed for most velocities and temperatures. 
%
At low electric field, friction decreases with load given the suppression of the interlocking and the increasing interaction between chains of the same layer.
For intermediate electric fields the load-dependence is not monotonic and for large field friction increases with load given the promotion of the interactions between chains of opposite layers.

While the specific quantitative detail of these results is to be traced to the peculiar electro-mechanical and geometric properties of the investigated model, and to the limitations intrinsic to modeling, we can generally conclude that friction of dipolar compounds can be altered even substantially by sufficiently large external DC electric fields. 
We argue that confined bi-layers 
can switch from smooth sliding to stick-slip regimes in a complex and non trivial, thus fascinating, way.
Compared to the slow thermal diffusion of free ions, dipole reorientation has the advantage of a faster response, mostly limited by molecular inertia: this observation suggests that the zwitterionic bi-layer 
approach may lead to a sizable response in the AC regime, especially at frequencies exceeding $\sim 10~\text{kHz}$, where ion diffusion usually becomes negligible.

\section*{Author Contributions}

MMG, RG and NM performed the numerical investigation and the data analysis. 
All authors discussed and collaborated to the formal analysis. 
All authors participated in the writing and revisions of this paper.

\section*{Conflicts of interest}

There are no conflicts to declare.

\begin{acknowledgement}

The authors acknowledge support from the grant PRIN2017 UTFROM of the Italian Ministry of University and Research 
and from the University of Milan through the APC initiative.
A.V.\ acknowledges also support by ERC Advanced Grant ULTRADISS, contract No. 8344023.
M.U.\ acknowledges the financial support of the Israel Science Foundation, Grant 1141/18.
The authors acknowledge useful discussions with Di Jin, Jacob Klein, Erio Tosatti and Yu Zhang. 
\end{acknowledgement}

\begin{suppinfo}


Snapshot of a simulation illustrating the fluid-like 
arrangement at high fields at room temperature (Figure S1);
average shear stress as a function of $v_\text{stage}$ and 
the energetics of the system at $T = 0~\text{K}$ and 
$E = 4~\text{V}\cdot\text{nm}^{-1}$ (Figure S2); 
average shear stress as a function of $v_\text{stage}$ and 
the energetics of the system at $T = 300~\text{K}$ and 
$E = 5~\text{V}\cdot\text{nm}^{-1}$ (Figure S3); 
average shear stress as a function of $v_\text{stage}$ and 
the energetics of the system at $T = 300~\text{K}$ and 
$E = 10~\text{V}\cdot\text{nm}^{-1}$ (Figure S4); 
average shear stress at room temperature and 
$E = 10~\text{V}\cdot\text{nm}^{-1}$ and hooking fraction 
as a function of load at room temperature for 
$E =$ 1, 5 and $10~\text{V}\cdot\text{nm}^{-1}$ (Figure S5).

Movie1: 3 ns of the MD simulation corresponding to the snapshots shown in Figure~\ref{snaps300_low}, $E = 2~\text{V}\cdot\text{nm}^{-1}$ at $T = 300~\text{K}$ (MP4).

Movie2: 3 ns of the MD simulation corresponding to the snapshots shown in Figure~\ref{snapsEF_5}, $E = 5~\text{V}\cdot\text{nm}^{-1}$ at $T = 300~\text{K}$ (MP4).

Movie3: 3 ns of the MD simulation corresponding to the snapshots shown in Figure~S1, and the friction trace in Figure~S4d,l, $E = 10~\text{V}\cdot\text{nm}^{-1}$, $T = 300~\text{K}$ (MP4) and $v_\text{stage}= 5~\text{m}\cdot\text{s}^{-1}$.

\end{suppinfo}

\providecommand{\latin}[1]{#1}
\makeatletter
\providecommand{\doi}
  {\begingroup\let\do\@makeother\dospecials
  \catcode`\{=1 \catcode`\}=2 \doi@aux}
\providecommand{\doi@aux}[1]{\endgroup\texttt{#1}}
\makeatother
\providecommand*\mcitethebibliography{\thebibliography}
\csname @ifundefined\endcsname{endmcitethebibliography}
  {\let\endmcitethebibliography\endthebibliography}{}

\end{document}





\begin{figure}[tb]
\includegraphics[width=0.5\textwidth]{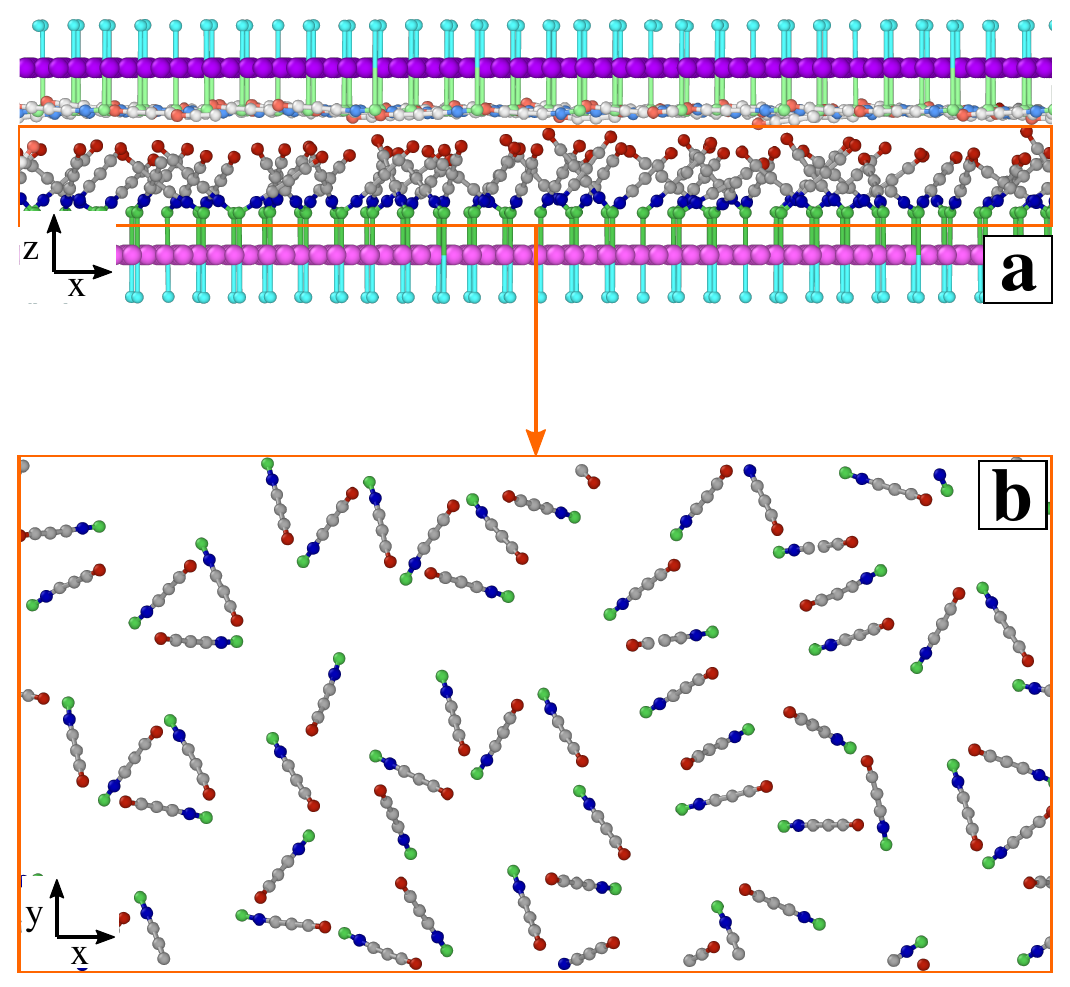}
\caption{
(a) Side view and (b) top view of a  4~nm $y$-thick slice of a snapshot of a simulation carried out with $E = 10~\text{V}\cdot\text{nm}^{-1}$, at $T = 300~\text{K}$. 
For better visibility, the top view includes the SUB chains only, i.e.\ the region inside the orange rectangle in panel (a).
\label{snapsEF_10}
}
\end{figure}

\begin{figure}[tb]
\includegraphics[width=0.8\textwidth]{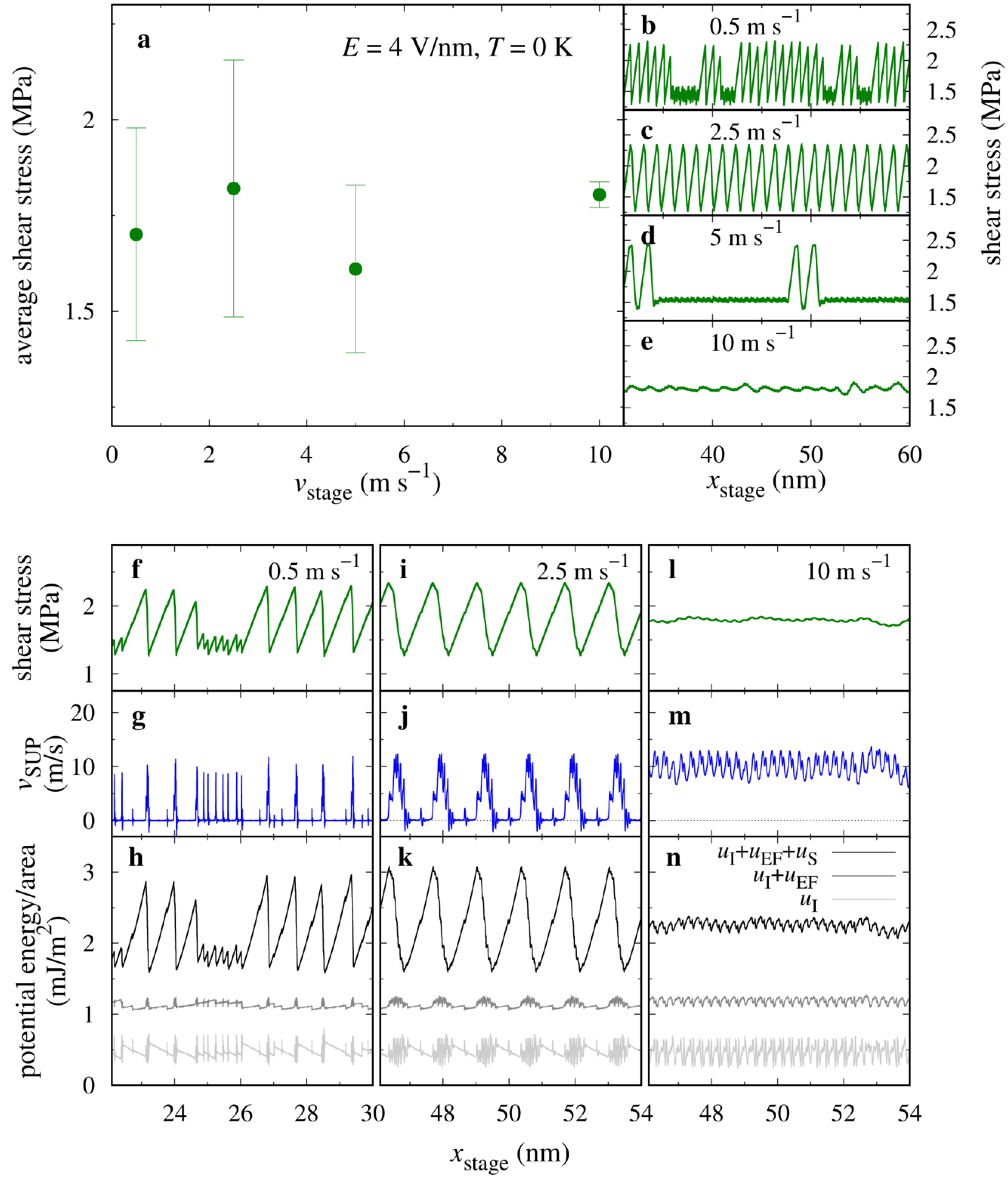}
\caption{\label{fig:vX_0_4} 
(a) Average shear stress as a function of $v_{\rm stage}$ for $E = 4~\text{V}\cdot\text{nm}^{-1}$, $T = 0~\text{K}$, and $L = 10~\text{MPa}$. 
(b-e) Shear-stress traces. 
Instantaneous
(f,i,l) shear stress, 
(g,j,m) velocity of the SUP layer, and 
(h,k,n) per unit area potential-energy contributions as a function of $x_{\rm stage}$ for the corresponding velocities in panel (a). 
$u_\text{I} = U_\text{I}/A$ is the internal potential energy; 
$u_\text{EF} = U_\text{EF}/A$ is the potential energy for the interaction with the applied electric field; 
$u_\text{S} = U_\text{S}/A$ is the potential energy of the pulling spring.
}
\end{figure}

\begin{figure}[b]
\includegraphics[width=0.8\textwidth]{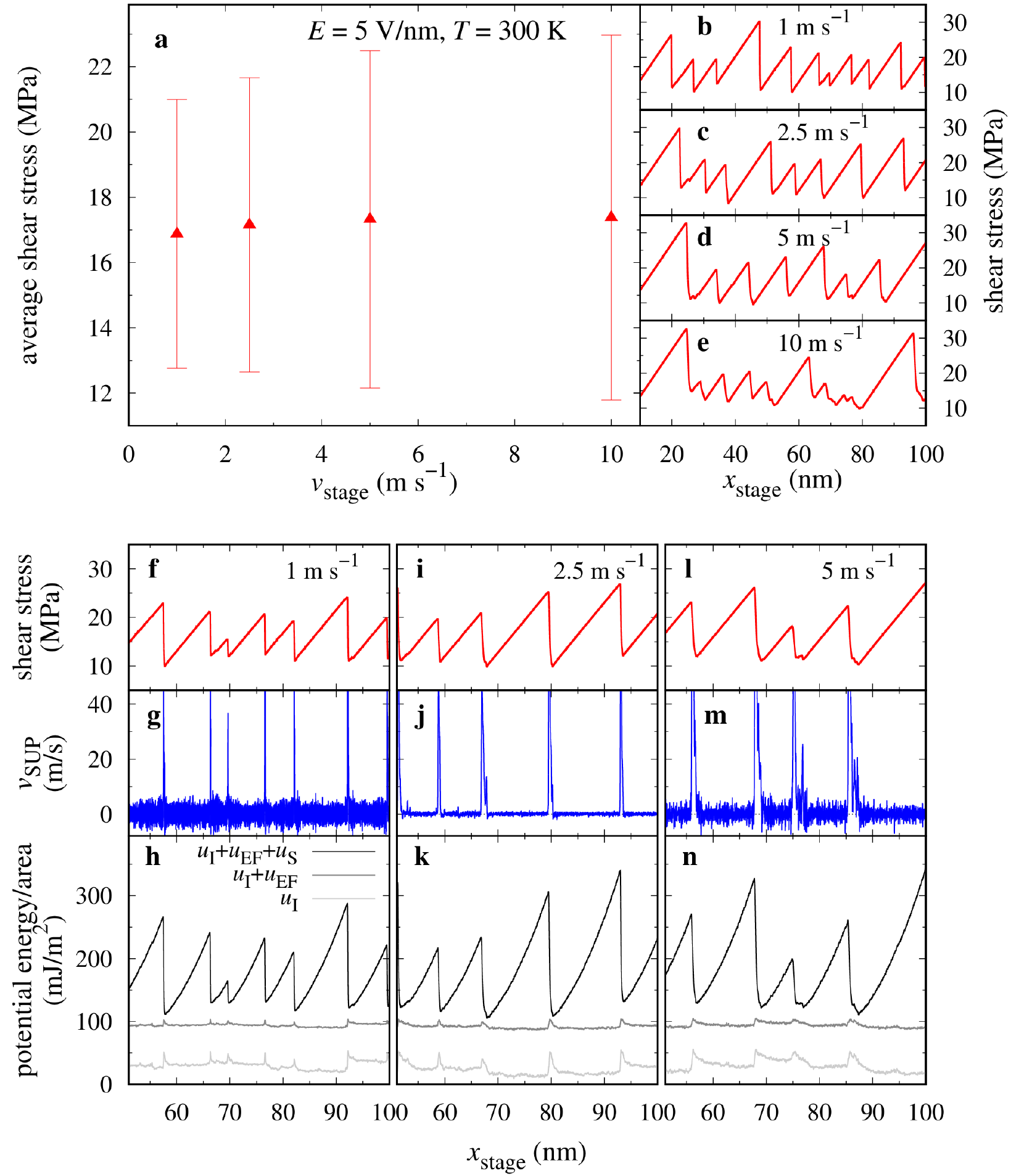}
\caption{\label{fig:vX_300_5}
  (a) Average shear stress as a function of $v_{\rm stage}$ for $E = 5~\text{V}\cdot\text{nm}^{-1}$, $T = 300~\text{K}$ and $L = 10~\text{MPa}$. 
  (b-e) Shear-stress traces for the velocities reported in panel (a). 
  (f,i,l) Shear stress, (g,j,m) instantaneous velocity of the SUP layer and 
  (h,k,n) per unit area potential energy contributions as a function of $x_{\rm stage}$ for the corresponding velocities in panel (a). 
  In the analysis of the potential energy contributions (panels h,k,n), thermal noise affects the data heavily: to discern readable energy signals we apply a Gaussian smoothing with width of 0.05~nm.
To mitigate the thermal noise in the SUP velocity we evaluate its average value by means of finite differences of the SUP position over the time corresponding to the driving stage advancing by 0.05~nm.
}
\end{figure}


\begin{figure}[tb]
\includegraphics[width=0.8\textwidth]{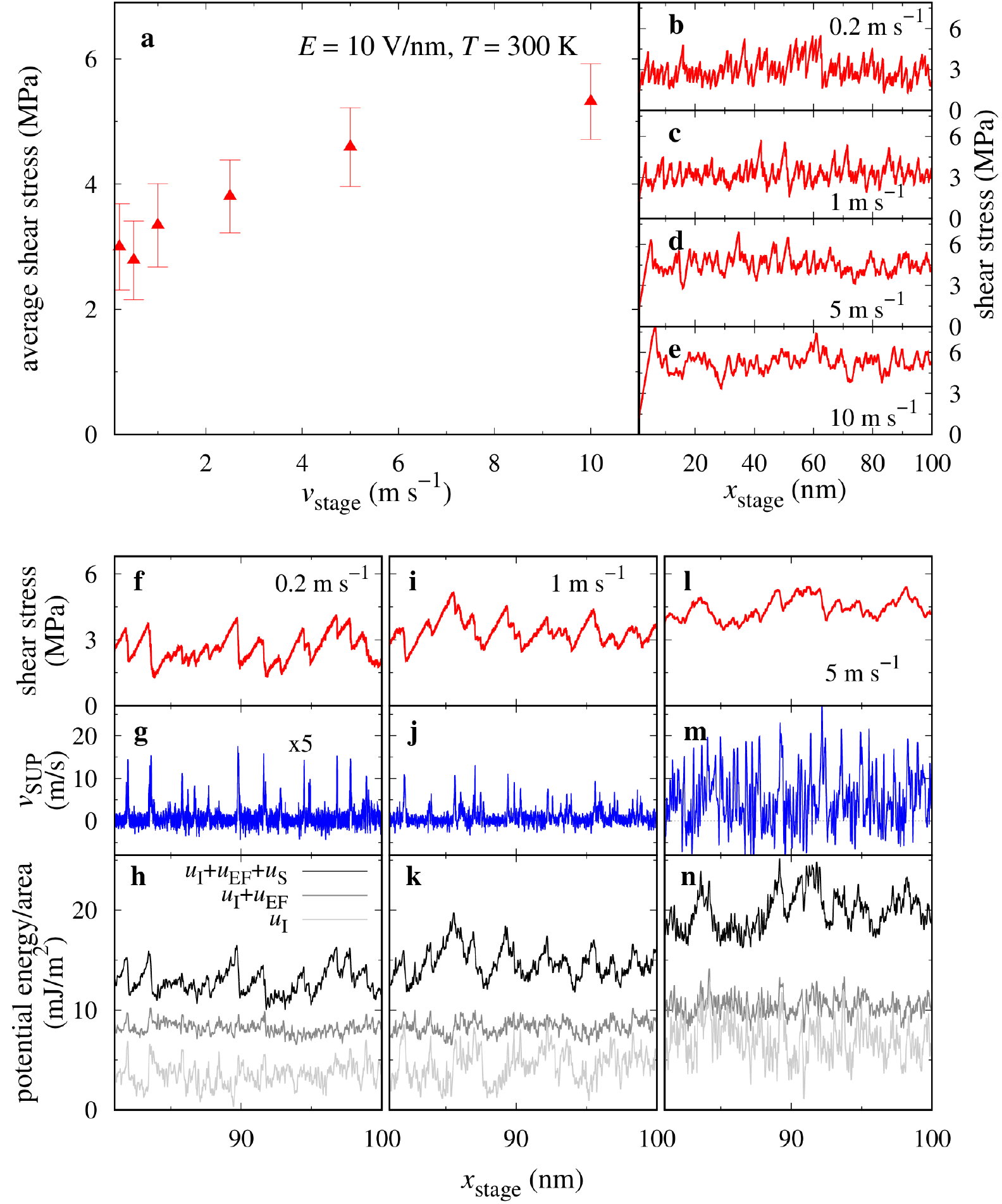}\\
\caption{
  (a) Average shear stress as a function of $v_{\rm stage}$ for $E = 10~\text{V}\cdot\text{nm}^{-1}$, $T = 300~\text{K}$ and $L = 10~\text{MPa}$. 
  (b-e) Shear traces for a few of the velocities reported in panel (a). 
  (f,i,l) Shear stress, 
  (g,j,m) instantaneous velocity of the SUP layer and 
  (h,k,n) per unit area potential energy contributions as a function of $x_{\rm stage}$ for the corresponding velocities in panel (a). 
  Thermal noise has been mitigated like in Figure~S3. 
  The intense electric field keeps the SUP layer of chains flat to the point of remaining nearly ``frozen'', thus preventing the entanglement of SUB chains. 
  Friction traces at the lowest velocities do show some hints of stick points (Figure~S4f, i) and eventually at low speed friction seems to depend only weakly on the sliding velocity, Figure~S4a.
Stick-slip features are visible in the SUP velocity and total potential energy, while they are harder to detect in the internal contribution $U_\text{I}$.
Even at $T=300$~K, signs of the unexpected slight decrease of $U_\text{I}$ during stick can be detected near $x_\text{stage}\simeq 82~\text{nm}$ and $x_\text{stage}\simeq 90~\text{nm}$, similar to that discussed for $T=0$ in the main text.
}
\label{fig:vX_300_10}
\end{figure}

\begin{figure}[b]
\includegraphics[width=0.8\textwidth]{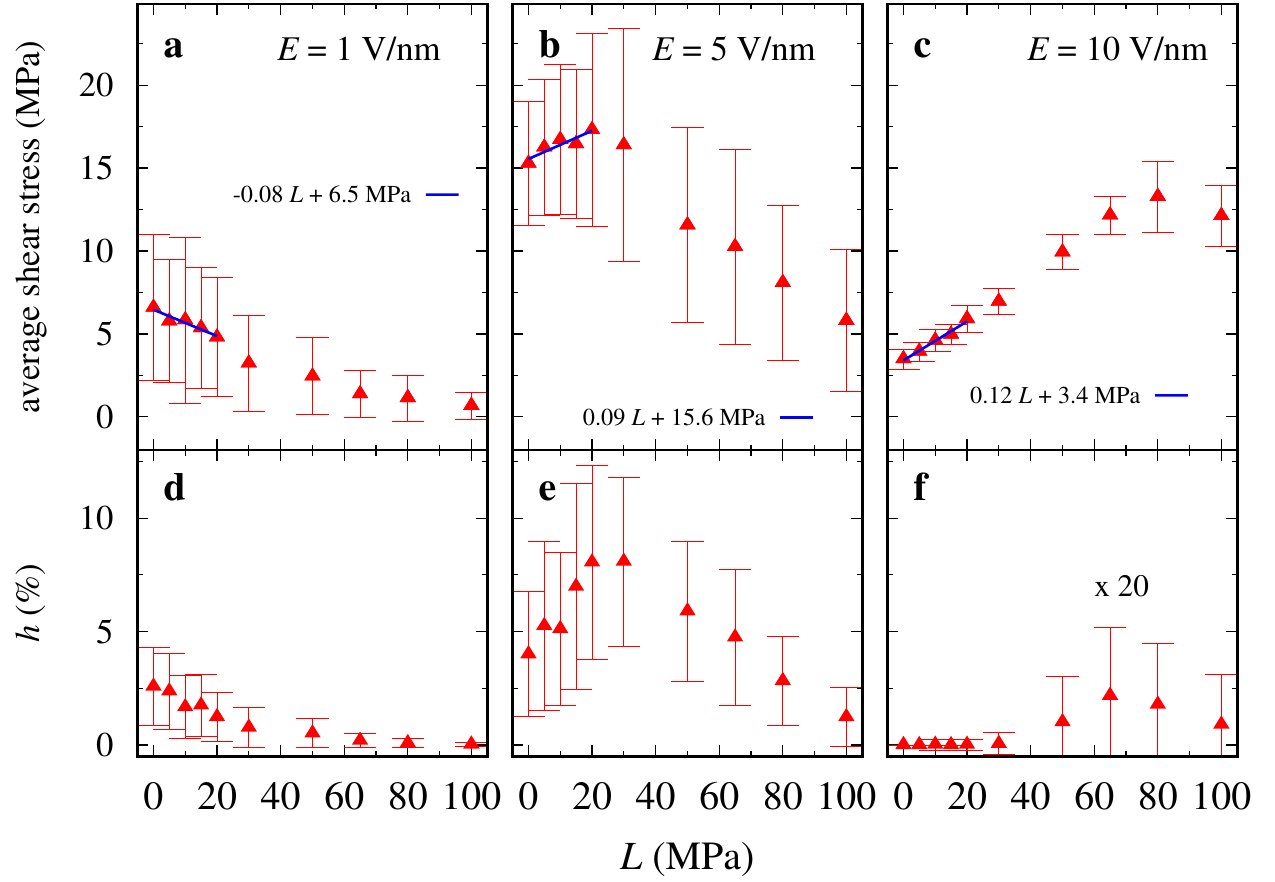}
\caption{\label{fig:hook_300}
  (a-c) Average shear stress (same as Fig.~7 of the article) and 
  (d-e) average hooking fraction $h$ (our strategy for quantifying the degree of interlocking defined in the Supporting Information of Ref.~\citenum{gianetti2022}) as a function of the applied load $L$ at $T = 300~\text{K}$.
  Blue lines are linear fits of the small-load (up to 20~MPa) data.
The resulting differential friction coefficients are:
$\mu_d = -0.08$ for $E =  1~\text{V}\cdot\text{nm}^{-1}$;
$\mu_d = 0.09$ for $E =  5~\text{V}\cdot\text{nm}^{-1}$;
$\mu_d = 0.12$ for $E = 10~\text{V}\cdot\text{nm}^{-1}$.
}
\end{figure}
\clearpage\newpage
\section{SI Movies}

Each of the SI movies reports 3~ns (corresponding to a $15$~nm advancement of the stage) 
of a MD simulation.
In simulation time, the frame rate is 1 frame every 20~ps. 
In running time, the frame rate is 10 frames per second. 
For clarity, like in Fig.~S1, 
the movies only include a 5 nm $y$-thick slice of the simulation cell 
(whose entire $y$-side is 14.41~nm). 
One of the particles of the SUP layer is drawn of bigger size 
and lighter color to improve the visibility of the advancement of this layer.

\begin{itemize}
    \item \textbf{Movie1.mp4}:
    3~ns of a 
    $E = 2~\text{V}\cdot\text{nm}^{-1}$, $v_\text{stage}= 5~\text{m}\cdot\text{s}^{-1}$, $T = 300~\text{K}$ simulation, also reported in the snapshots of Figure~3 of the main text;
    \item \textbf{Movie2.mp4}: 3~ns of a 
    $E = 5~\text{V}\cdot\text{nm}^{-1}$, $v_\text{stage}= 5~\text{m}\cdot\text{s}^{-1}$, $T = 300~\text{K}$ 
    simulation, also reported in the snapshots of Figure~4b,c,e,f 
    of the main text; 
    \item \textbf{Movie3.mp4}:
    3~ns of a 
    $E = 10~\text{V}\cdot\text{nm}^{-1}$, $v_\text{stage}= 5~\text{m}\cdot\text{s}^{-1}$, $T = 300~\text{K}$ 
    simulation, also reported in the snapshots shown in Figure~S1 and in the friction trace shown in Figure~S4d,l of the present SI.
    \end{itemize}
\clearpage\newpage


\providecommand{\latin}[1]{#1}
\makeatletter
\providecommand{\doi}
  {\begingroup\let\do\@makeother\dospecials
  \catcode`\{=1 \catcode`\}=2 \doi@aux}
\providecommand{\doi@aux}[1]{\endgroup\texttt{#1}}
\makeatother
\providecommand*\mcitethebibliography{\thebibliography}
\csname @ifundefined\endcsname{endmcitethebibliography}
  {\let\endmcitethebibliography\endthebibliography}{}